\documentclass{moriond}

\def\be{\begin{equation}}
\def\ee{\end{equation}}
\def\bea{\begin{eqnarray}}
\def\eea{\end{eqnarray}}

\newcommand{\Photo}{\includegraphics[height=35mm]{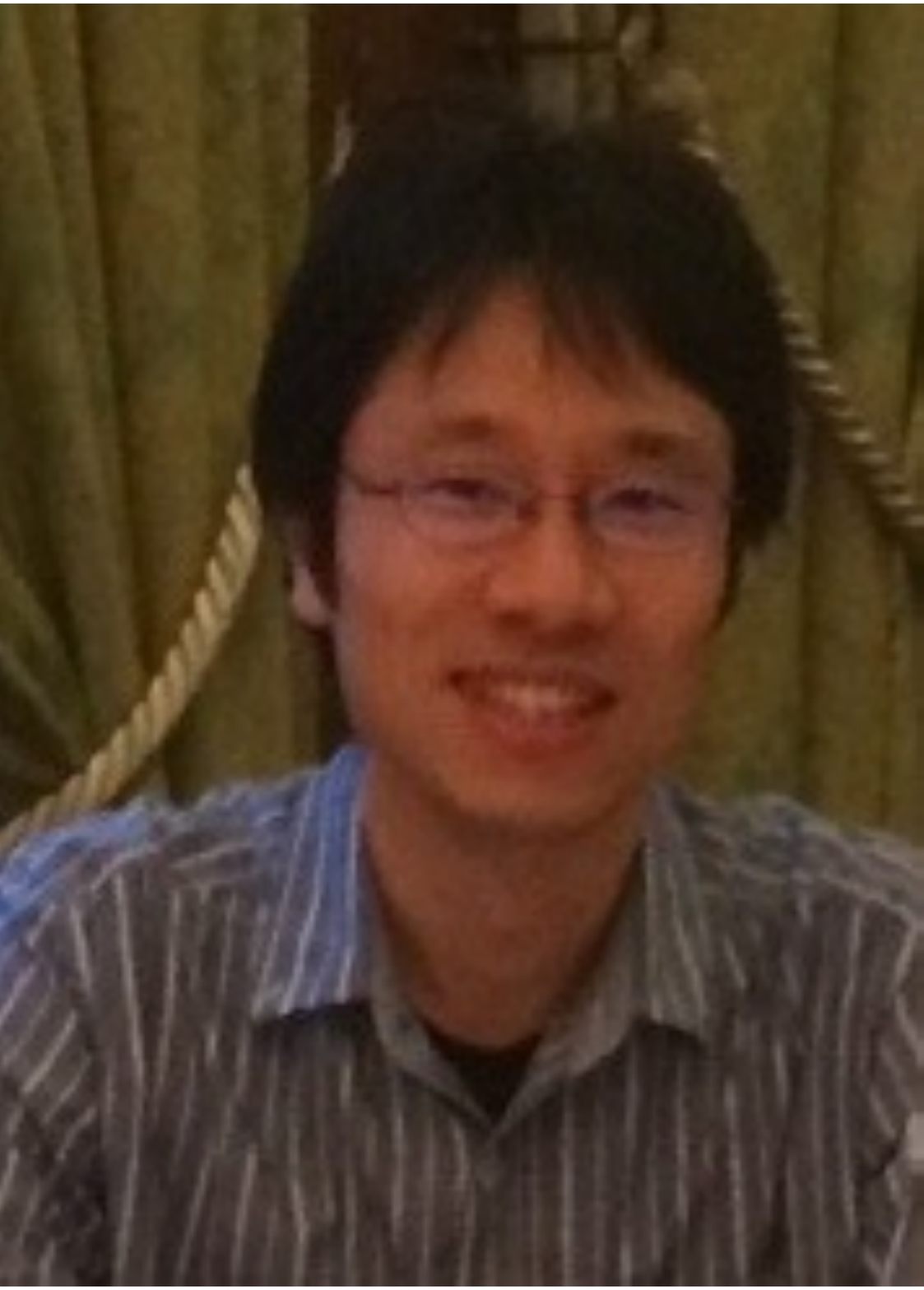}}
\newcommand{\mLQ}{m_\mathrm{LQ}}
\newcommand{\tauh}{\tau_\mathrm{h}}

\newcommand{\etau}{\mathrm{e}\tau_\mathrm{h}}
\newcommand{\mutau}{\mu\tau_\mathrm{h}}
\newcommand{\tautau}{\tau_\mathrm{h}\tau_\mathrm{h}}
\newcommand{\ltau}{\ell \tau_\mathrm{h}}
\newcommand{\ptmiss}{p_\mathrm{T}^\mathrm{miss}}
\newcommand{\pt}{p_\mathrm{T}}
\newcommand{\HT}{H_\mathrm{T}}
\newcommand{\mt}{M_\mathrm{T2}}
\newcommand{\nj}{N_\mathrm{j}}
\newcommand{\ttbar}{\mathrm{t}\bar{\mathrm{t}}}
\newcommand{\ptt}{\pt^\mathrm{t}}

\begin{document}
\vspace*{4cm}
\title{LEPTOQUARK SEARCHES IN CMS}

\author{ YUTA TAKAHASHI \\on behalf of the CMS Collaboration}

%\address{Universit\"{a}tZ\"{u}rich, Physik-Institut, Winterthurerstrasse 190, CH-8057 Z\"{u}rich, Switzerland}
\address{Universit\"{a}t Z\"{u}rich, Physik-Institut, Winterthurerstrasse 190, CH-8057 Z\"{u}rich, Switzerland}

\maketitle

\abstracts{
Leptoquarks (LQs) are hypothetical bosons, each coupling to a lepton and a quark of a given generation. 
They recently got particular attention, as the ``TeV-scale LQ" might explain 
observed anomalies reported by dedicated B physics experiments. 
In these proceedings, the recent progress of the direct searches on LQs at CMS experiment is presented.
The results are based on a data sample of proton-proton collisions at a center-of-mass energy of 13~TeV,
recorded with the CMS detector corresponding to an integrated luminosity of 35.9~fb$^{-1}$.
}

\section{Introduction}
Leptoquarks (LQs) are hypothetical color-triplet bosons, that carry both baryon and lepton
quantum numbers, and have fractional electric charge. 
They are predicted by many extensions of the standard model (SM) of particle physics, such as grand unified theories, technicolor frameworks, 
and composite models. Due to experimental constraints on flavor
changing neutral currents and other rare processes, it is generally assumed that there
would be three types of LQs, each coupled to leptons and quarks of the same generation.
The second- and third-generation LQs have recently received considerable theoretical interest, 
as the TeV scale LQ might provide explanations for the B physics anomalies reported by the
BaBar, Belle, and LHCb Collaborations. In the following sections, recent progress of the direct searches on LQs at CMS experiment~[1] is reviewed.

\section{LQ signatures}
The production cross sections in proton-proton colliders are determined by the mass of the LQ ($\mLQ$)
and the Yukawa coupling $\lambda$ of the LQ-lepton-quark vertex. 
LQs can be produced in pairs via gluon fusion or quark-antiquark annihilation, and singly via quark-gluon fusion. Pair production of
LQs does not depend on $\lambda$, while single production does, and thus the sensitivity of single-LQ searches depends on $\lambda$. 
At lower masses, the cross section for pair production is greater
than that for single production. However, single-LQ production cross section decreases more
slowly with increasing $\mLQ$, exceeding that for pair production with masses above $\mathcal{O}$(1) TeV.

The LQ decay depends on the (unknown) branching fraction to a charged lepton ($\ell$) and a quark (q), denoted as $\beta = \mathcal{B} (\mathrm{LQ} \to \ell \mathrm{q})$,
whereas the rest of the decay will be to a neutrino ($\nu$) and a quark; $1-\beta = \mathcal{B} (\mathrm{LQ} \to \nu \mathrm{q})$. 
Depending on the assumed $\beta$ value, the preferred decay signatures vary as summarized in Table~\ref{signature}. 
The recent search results are marked with a *, which will be reviewed in the rest of the proceedings.
\begin{table}[h!]
\caption{Possible decay signatures of the LQ. The results marked with a * are reviewed in these proceedings.}
\vspace{0.2cm}
\begin{center}
\begin{tabular}{|l|ccc|ccc|} \hline
LQ coupling & \multicolumn{3}{c}{Pair LQ production} & \multicolumn{3}{|c|}{Single LQ production} \\ \hline
 1st. generation & eejj~[2] & e$\nu$jj [2] & $\nu\nu$jj*~[5] & eej & e$\nu$j & $\nu\nu$j \\
 2nd. generation & $\mu\mu$jj*~[3] & $\mu\nu$jj*~[3] & $\nu\nu$jj*~[5] & $\mu\mu$j & $\mu\nu$j & $\nu\nu$j \\
 3rd. generation (q = t) & $\tau\tau $tt*~[4] & $\tau \nu$tt & $\nu\nu$tt*~[5] & $\tau\tau$t & $\tau \nu$t & $\nu\nu$t \\
 3rd. generation (q = b) & $\tau\tau $bb~[6,7] & $\tau \nu$bb & $\nu\nu$bb*~[5]  & $\tau\tau$b*~[8] & $\tau \nu$b & $\nu\nu$b \\ \hline
\end{tabular} 
\end{center} 
\label{signature}
\end{table} 

\section{LQLQ$\to \mu\mu$jj, $\mu\nu$jj}
The LQ couplings to the 2nd generation fermions are of importance for two reasons; 1) they might explain observed anomalies in B$\to$K$^{(*)}\mu^{+}\mu^{-}$ decay, 
and 2) the third-generation LQ, once produced, might result in final states including muons at the end of its decay chain. 
We searched for the final states with two muons 
and two jets ($\mu\mu$jj channel) or with one muon, two jets, and missing transverse energy ($\ptmiss$) ($\mu\nu$jj channel)~[3].
The $\mu\mu$jj ($\mu\nu$jj) channel drives the sensitivity at high (low) $\beta$, 
but the $\mu\mu$jj channel dominates the analysis sensitivity up to reasonably low $\beta$ region due to better S/B ratio. 
The description below focuses on the $\mu\mu$jj channel. 

The analysis requires two muons with transverse momentum, $\pt > 53$~GeV and pseudorapidity, $|\eta| < 2.4$, which are required to satisfy a set of identification criteria optimized for high $\pt$.
%An isolation requirement is imposed to select high-quality muons, where the sum of the transverse momenta of all tracks in the tracker originating from the primary vertex in a cone of $\Delta R =  \sqrt{\Delta \phi^2 + \Delta \eta^2} = 0.3 $ around the muon track (excluding the muon track itself), divided by the muon $p_T$, is less than 0.1.
In addition, at least two jets are required, each should satisfy $\pt > 50$~GeV, $|\eta| < 2.4$, and to be separated from selected muons. % by $\Delta R = 0.5$.
The LQ candidates are then reconstructed by pairing each muon with a jet in the configuration that minimizes the LQ-$\overline{\mathrm{LQ}}$ invariant mass difference.

The signal extraction is performed by a cut and count analysis. For each mass point, an optimized set of cuts for three kinematic variables is applied: $S_\mathrm{T}^{\mu\mu \mathrm{jj} }$, $M_{\mu\mu}$, and $M^{\mathrm{min}}_{\mu j}$, 
where $S_\mathrm{T}^{\mu\mu \mathrm{jj}}$ is the scalar sum of the transverse momenta of the two jets and two muons in the event, 
$M_{\mu\mu}$ is the invariant mass of the di-muon system, and 
$M^{\mathrm{min}}_{\mu j}$ is defined as the smaller of the two muon-jet invariant masses that minimize the LQ-$\overline{\mathrm{LQ}}$ invariant mass difference.
%A full three-dimensional optimization is performed, with signal-to-background separation optimized using the Punzi significance,
%which is optimal for both setting limits and for making a discovery, and is valid in cases with low background statistics.
At high $\mLQ$, the harder cuts are applied, as it helps to effectively reduce backgrounds. % while maintaining majority of the signal events. 

\begin{figure}[h!]
\begin{center}
\vspace{-2cm}
\includegraphics[height=13cm,angle=90]{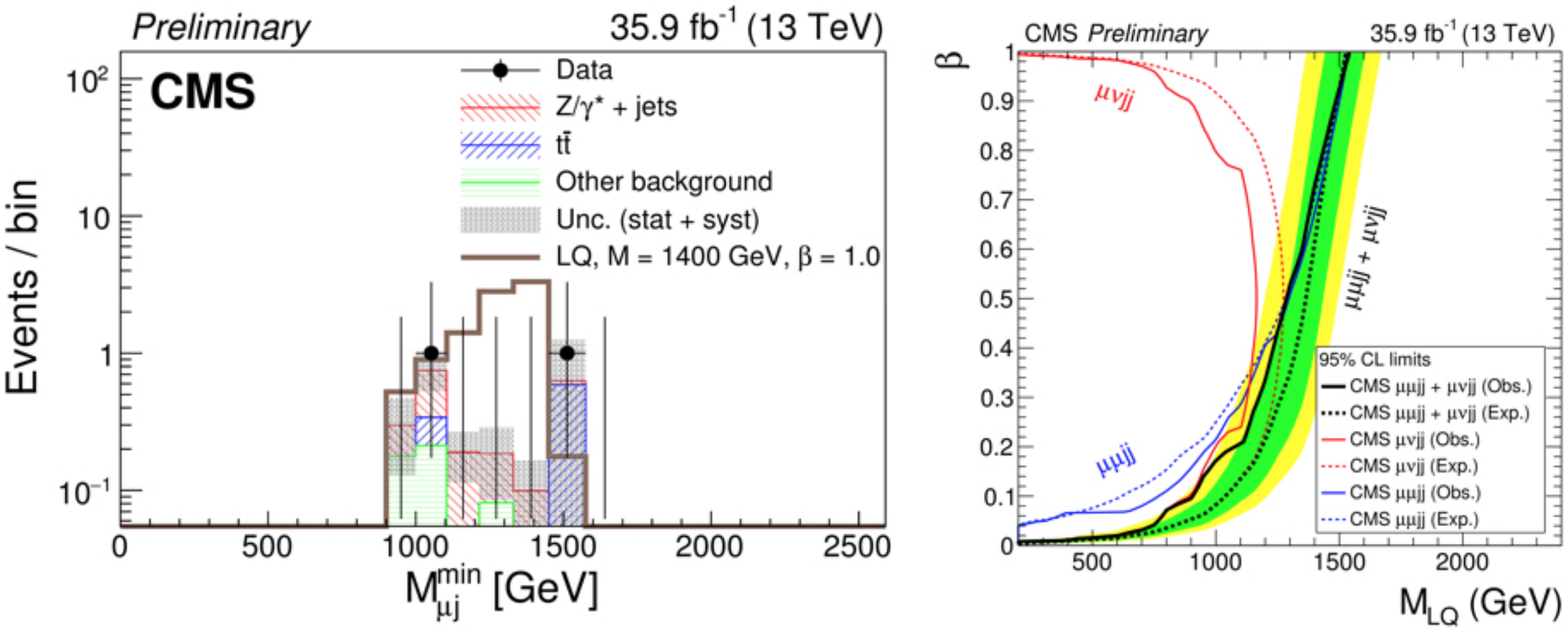}
\end{center}
\vspace{-2.5cm}
\caption{(Left) The $M^{\mathrm{min}}_{\mu j}$ distribution at final selection level for $\mLQ = 1400$~GeV for the $\mu\mu$jj channel, (right) the expected and observed exclusion limits at 95\% CL for second-generation LQ mass as a function of $\beta$ [3].}
\label{fig:1}
\end{figure}

Figure~\ref{fig:1} (left) shows the $M^{\mathrm{min}}_{\mu j}$ distribution for the mass point of $\mLQ = 1400$~GeV. 
The dominant backgrounds are Z$+$jets, $\ttbar$ and the single top quark processes, which are estimated in a data-driven way, wherever possible. 
The analysis sensitivity is ultimately limited by the size of the data sample.
No significant excess is observed across the whole $\mLQ$ range considered in the analysis. 
Limits are set at 95\% confidence level (CL) for $\beta$ values from 0 to 1, as a function of $\mLQ$, as also shown in Figure~\ref{fig:1} (right). The combination improves the mass exclusion for values of $\beta < 1$. Second-generation scalar LQs with $\mLQ < $ 1530 (1285) GeV are excluded for $\beta$ = 1.0 (0.5). 

\section{LQLQ$\to \nu\nu$jj, $\nu\nu$bb , $\nu\nu$tt}
The results of a previous search~[9] for squarks and gluinos
are recently reinterpreted~[5] to constrain models of scalar and vector LQ production. The LQ decays to a neutrino and a top, bottom, or light quark (any single one of up, down, strange, or charm) are considered.
This reinterpretation relies on our confirmation that the acceptance of the analysis does not change, within uncertainties, between squark, scalar LQ,
and vector LQ pair production for the same particle mass.
We extended the interpretations to higher mass values, incorporating higher cross sections of the vector LQ signals.
%production cross sections in the $t\nu$ decay channel by as much as a factor of 2.8 over the flat extrapolation assumed in Ref. [36].

The baseline selection requires $\nj \geq 1$, where $\nj$ denotes the number of jets with $\pt > 30$~GeV and $|\eta| < 2.4$,
and to pass either $\ptmiss > 30$~GeV if they have $\HT > 1000$~GeV, or $\ptmiss > 250$~GeV if they have $250 < \HT < 1000$~GeV,
where $\HT$ is defined as the scalar sum of jet $\pt$.
Further baseline requirements include that the $\ptmiss$ vector is not aligned in the azimuthal angle $\phi$ with any of the four leading jets in $\pt$, 
that the negative vector sum of jet transverse momenta is consistent with the $\ptmiss$ vector, and that no loosely identified charged leptons or isolated tracks are found in the event.

For events with $\nj \geq 2$, 
a momentum imbalance, $\mt > 200$~GeV is required, which is raised to $\mt > 400$~GeV for events with $\HT > 1500$~GeV, to further reject QCD multijet backgrounds.
%Here, $\mt$ is computed from the jets and the $\ptmiss$ vector and takes small values for events where the momentum imbalance arises from jet mismeasurement (e.g. QCD multijet background)
%and yields larger values in events with genuine $\ptmiss$.
Events are then categorized according to four variables:
$\HT$, $\mt$, $\nj$, and number of b-tagged jets. 
%Here,  the variable $\mt$ is computed from
%the jets and the $\ptmiss$ vector as described in Ref. [46]. 
%The $\mt$ variable is a measure of momentum imbalance, which takes small values for events where the momentum imbalance arises from jet mismeasurement (e.g. QCD multijet
%background) and yields larger values in events with genuine $\ptmiss$. 
Events with $N_j = 1$ are categorized according to the jet $\pt$ and the
presence or absence of a b-tagged jet. The analysis spans a wide range of kinematics and jet
multiplicities, containing 213 search bins in total. 
Depending on $\mLQ$ and the decay products, different search bins provide the greatest
signal sensitivity. 

\begin{figure}[h!]
\begin{center}
\vspace{-2cm}
\includegraphics[height=13cm, angle=90]{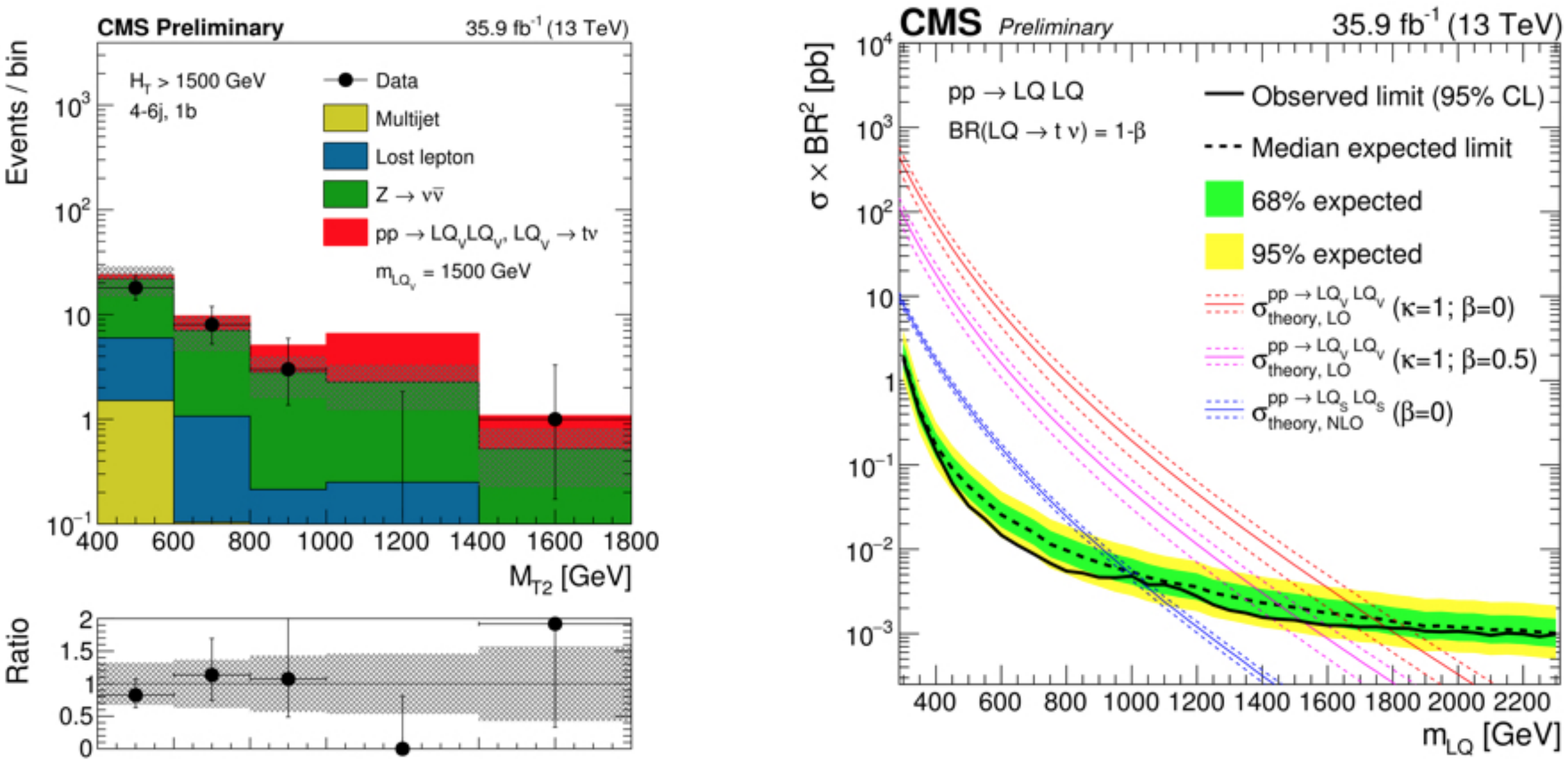}
\end{center}
\vspace{-2cm}
\caption{(Left) $\mt$ distribution with a hypothetical vector LQ signal with $\mLQ = 1500$~GeV with unity branching fraction to t$\nu$, (right) The 95\% CL upper limits on the production cross sections as a function of LQ mass for LQ pair production decaying with unity branching fraction to a neutrino and a top quark~[5].}
\label{fig:2}
\end{figure}

Figure~\ref{fig:2} (left) shows the $\mt$ distribution in one of the most sensitive search categories for $\mLQ = 1500$~GeV decaying with unity branching fraction to t$\nu$.
Simultaneous maximum likelihood fits to data yields in all bins are performed,  showing no significant deviations from the SM prediction.
Assuming a scalar (vector) LQ decaying with unity branching fraction to a light quark and neutrino, $\mLQ < $ 980 (1790) GeV are excluded at the 95\% CL by the observed data. For an LQ decaying to b$\nu$, $\mLQ < $ 1100 (1810) GeV are excluded, and for an LQ decaying to t$\nu$, $\mLQ < $ 1020 (1780) GeV are excluded, as shown in Figure~\ref{fig:2} (right). 
A vector LQ decaying with a 50\% branching fraction to t$\nu$, and 50\% to b$\tau$, has been proposed~[10] as part of an explanation of anomalous B physics results. In such a model, using only the decays to t$\nu$, LQ masses below 1530~GeV are excluded assuming the Yang-Mills type coupling, placing the most stringent constraint to date from pair production.

\section{LQLQ$\to \tau\tau$tt}
This analysis~[4] targets pair production of LQs, each decaying to a top quark and a $\tau$ lepton. Since both the top quarks and the $\tau$ leptons
can decay leptonically or hadronically, a variety of final states emerges.
The search considers final states with an electron ($\pt > 30$~GeV, $|\eta| < 2.1$) or a muon ($\pt > 30$~GeV, $|\eta| < 2.4$), one or two $\tauh$ candidate(s) ($\pt > 20$~GeV and $|\eta| < 2.1$), 
and at least two jets ($\pt > 50$~GeV and $|\eta| < 2.4$).
Events are selected if a third jet with $\pt > 30$~GeV and
$|\eta| < 2.4$ is present, and any additional jets are only considered if they have $\pt > 30$~GeV. 
The events are further divided into three categories, depending on various kinematic quantities. 
The most sensitive category requires 
exactly one $\tauh$ candidate with $\pt > 100$~GeV and one oppositely charged electron or muon. 
Further, the leading jet is required to have $\pt > 150$~GeV, $\ptmiss > 100$~GeV, at least one b-tagged jet,
and the scalar $\pt$ sum of all selected jets, leptons, and $\ptmiss$ should be greater than 1200~GeV. 

In this analysis, the dominant background comes from $\ttbar$ production process, where the jet is misidentified as a $\tauh$ candidate.
As the top quarks originating from the decay of a heavy LQ are expected to be produced with larger $\pt$ than the top quarks in the $\ttbar$ background, 
the $\pt$ distribution of the top quark candidate decaying into hadronic jets ($\ptt$) is used as a final discriminant. 
The $\ptt$ is constructed by the jet triplet that gives the best top quark mass (172.5~GeV).
Since the $\ptt$ distribution is not well modeled by the simulated events, the background distribution in the signal region (SR) has been derived 
by using a control region (CR), where the $\tauh$ isolation requirement is inverted. 
%The shape of the $\ptt$ distribution is compared between the CR and SR in simulated events. 
Since the inversion of the $\tauh$ isolation criterion introduces kinematic differences between the SRs and CRs, the $\ptt$ distribution has been corrected in order to reproduce the shape of the backgrounds in the SRs. 

\begin{figure}[h!]
\begin{center}
\vspace{-2cm}
\includegraphics[height=13cm, angle=90]{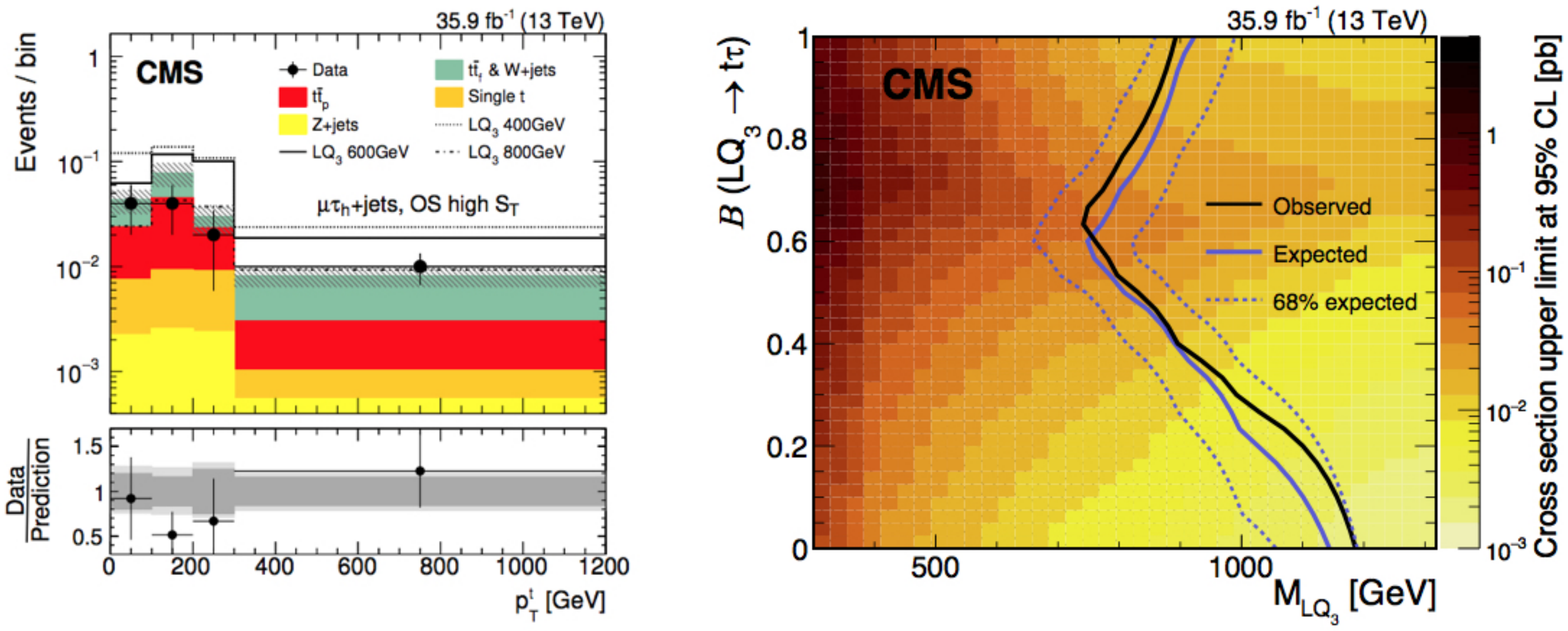}
\end{center}
\vspace{-2.5cm}
\caption{(Left) $\ptt$ distribution in one of the most sensitive category in the muon channel, (right) Upper limits at 95\% CL on the leptoquark mass as a function of the branching fraction 
for the pair production of scalar LQs decaying to a top quark and a $\tau$ lepton~[4].}
\label{fig:3}
\end{figure}

%The results of all search categories in the electron and muon channels are combined in a binned-likelihood fit. 
The statistical evaluation in the most-sensitive category is performed through a template-based fit to the measured $\ptt$ distribution, as shown in Figure~\ref{fig:3} (left) for the muon channel.
%The post-fit $\ptt$ distributions in the muon channel is shown in Figure~\ref{fig:3}.
No evidence for pair production of LQs is found. Assuming a branching fraction of unity for the decay LQ $\to \mathrm{t}\tau$, upper limits on the production cross section are set as a function of $\mLQ$, excluding masses below 900 GeV at 95\% CL. Exclusion limits with varying $\beta$ are presented in Figure~\ref{fig:3} (right), where limits on the complementary LQ $\to \mathrm{b}\nu$ ($\beta = 0$) decay channel are also included. The results for $\beta = 0$ are obtained from a search for pair-produced bottom squarks with subsequent decays into b quark and neutralino pairs~[9], in the limit of vanishing neutralino masses. Scalar LQs can be excluded for $\mLQ < 1200$~GeV for $\beta = 0$ and for $\mLQ < 750$~GeV over the full $\beta$ range. %Note that if upper limits on B are to be used to constrain the lepton-quark-LQ Yukawa couplings, $\lambda_{b\nu}$ and $\lambda_{t\tau}$, kinematic suppression factors that favor $b\nu$ decay over the $t\tau$ decay have to be considered as well.

\section{LQ$\to \tau \tau $b}
As the LQ mass increases, the cross section for the singly produced LQ starts to dominate than that of the pair production. 
If the third-generation LQ is responsible for the observed B physics anomalies, a large $\lambda$ is
favored ($\lambda \sim \mLQ$ measured in TeV), such that this crossover occurs at $\mLQ$ $= 1.0-1.5$~TeV~[10].
%Given the current limit from the LQ pair production, around 1TeV, search for singly produced LQ becomes more important. 
We searched for associated production of a LQ and a $\tau$ lepton, leading to a final state with a bottom quark and two $\tau$ leptons~[8].
The analysis selects events containing an electron or muon, and a $\tauh$ candidate (denoted as the $\etau$, $\mutau$ channel, respectively, or collectively as the $\ltau$
channels) or two $\tauh$ candidates ($\tautau$ channel), produced in association with a b-tagged jet. 

In the $\etau$ ($\mutau$) channel, events are selected by requiring a single electron (muon) with $\pt > 50$~GeV and $|\eta| < 2.1$ ($|\eta| < 2.4$), and a single $\tauh$ candidate with $\pt > 50$~GeV and $|\eta| < 2.3$. In the $\tautau$
channel, two $\tauh$ candidates are required with $\pt > 50$~GeV and $|\eta| < 2.1$. The selected lepton (electron or muon) and
$\tauh$ candidate, or two $\tauh$ candidates, must meet isolation requirements, and have opposite-sign electric charges. 
Since signal events contain at least one energetic bottom quark jet coming from the LQ decay, at least one b-tagged jet with
$\pt > 50$~GeV is required. To reduce the Z+jets background, the invariant mass,
$m_\mathrm{vis}$, of the lepton and $\tauh$ candidate (two $\tauh$ candidates), is required to be greater than 85 (95) GeV in the $\ltau$ ($\tautau$) channels.
The sensitivity of the analysis is dominated by the $\tautau$ channel due to its larger branching fraction of $\mathcal{B}(\tau\tau \to \tauh\tauh) = 42\%$, compared to $\mathcal{B}(\tau\tau \to \mu\tauh) = \mathcal{B}(\tau\tau \to e\tauh) = 21\%$.
Furthermore, $\ltau$ channels are contaminated by the $\ttbar \to \mathrm{WWbb} \to \ell\tauh \nu\nu \mathrm{bb}$, in addition to the $\ttbar \to \mathrm{WWbb} \to \tau_\ell \tauh \nu \nu \mathrm{bb} $ background, which is not the case for the $\tautau$ channel. Here, $\tau_\ell$ denotes a leptonically-decaying $\tau$ lepton. The dominant background in the $\tautau$ channel comes from QCD multijet and $\ttbar$ processes, where the former is derived in a data-driven way, and the latter by simulated events. 

%The efficiency for the singly produced LQ signals ranges from 0.2 to 1.3\% for $\mLQ$ between 200
%and 1500 GeV in the $\etau$ channel, including the branching fraction of $\tautau$ $\to$ $\etau$. The efficiency
%increases with increasing $\mLQ$ due to the harder $\pt$ spectra of the final state objects. Beyond
%1000 GeV, however, the acceptance starts to degrade mainly due to lower b tagging efficiencies.
%Similarly, the efficiencies in the $\mutau$ ($\tautau$) channels range from 0.3 to 1.8\% (0.5 to 2.5\%).

After applying the event selection, an excess of events over the SM backgrounds is searched for
using the distribution of the scalar $\pt$ sum of all required final state objects, $S_\mathrm{T}$.
%, which is defined
%as $\pt$(lepton) $+ \pt$($\tauh$) $+ \pt$(leading jet) for the $\ltau$ channels, and $\pt$(leading $\tauh$) $+ \pt$(subleading $\tauh$) $+
%\pt$(leading jet) for the $\tautau$ channel, where leading and subleading refer to $\pt$. 
A binned maximum likelihood method is used for the signal extraction. 
The fit is performed simultaneously in the $\etau$, $\mutau$, and $\tautau$ signal regions, 
as well as in the $e\mu$ control region to better constrain systematic uncertainties related to the $\ttbar$ modeling. 

\begin{figure}[h!]
\begin{center}
\vspace{-2cm}
\includegraphics[height=13cm, angle=90]{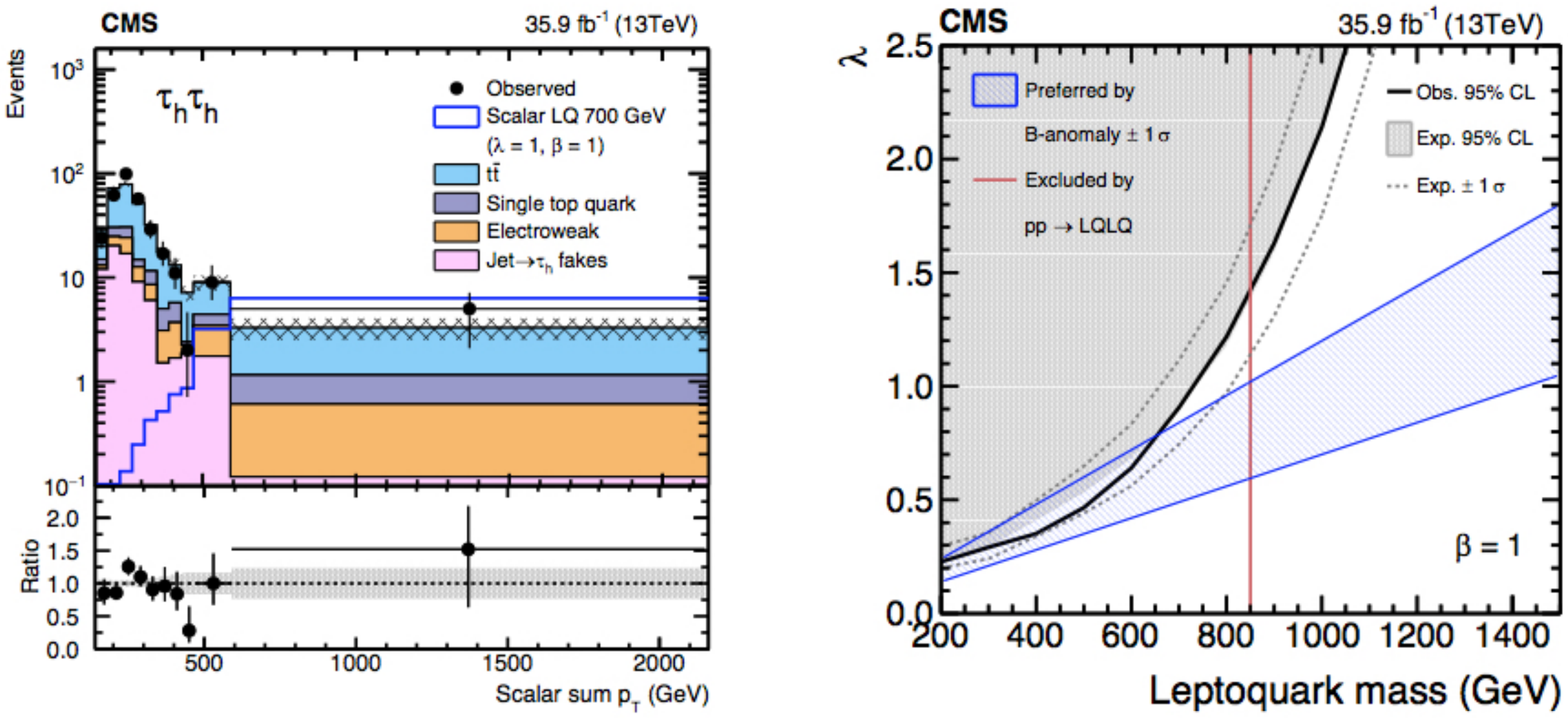}
\end{center}
\vspace{-2cm}
\caption{(Left) Scalar $\pt$ sum distribution in the $\tautau$ channel, (right) 95\% CL expected and observed exclusion limits on the Yukawa coupling $\lambda$
at the LQ-lepton-quark vertex, as a function of the LQ mass~[8].}
\label{fig:4}
\end{figure}

Figure~\ref{fig:4} (left) shows the $S_\mathrm{T}$ distributions in the $\tautau$ channel, after the combined fit. % to the $\etau$, $\mutau$, and $\tautau$ signal regions,
%as well as the $e\mu$ control region. The uncertainty bands on simulated event histograms represent the sum in quadrature of statistical and systematic uncertainties taking the full covariance
%matrix of all nuisance parameters into account. The dominant uncertainty in the background
%estimate comes from limited event count in simulated samples. However, uncertainties related
%to the simulated events play a small role for $\mLQ > 500$~GeV, where the limit of this analysis lies, and the measurement is ultimately limited by the size of the data sample.
The observed data are consistent with the background only (SM) hypothesis.
Assuming $\lambda = 1$ and $\beta = 1$, third-generation scalar LQs with $\mLQ$ below 750 (740) GeV are excluded at the expected (observed) 95\% CL.
Figure~\ref{fig:4} (right) also shows the expected and observed upper limits on $\lambda$ versus $\mLQ$. 
The blue band shows the parameter space for the scalar LQ preferred by the B physics anomalies:
$\lambda = 0.95 \pm 0.25 \times \mLQ$ (TeV)~[10]. The plot contains the limit from the pair production
search~[6], overlaid as a red vertical line, which does not depend on $\lambda$. 
This result, together with pair production search, begins to constrain the region of parameter space implied by the B physics anomalies.

\section{Summary}
CMS has searched for leptoquarks (LQs) as one of the primary benchmark scenarios with the lepton plus jets signature. 
In light of B physics anomalies, we performed extensive searches, especially in terms of second- and third-generation LQs. 
In these proceedings, recent progress has been highlighted: update on the LQ pair production decaying to muon final states;
reinterpretation of the squark, gluino searches in the context of scalar and vector LQs; update on the LQ pair production decaying to $\tau\tau$tt; new searches on singly-produced LQs decaying to $\tau$b. 
All results are consistent with SM expectations, and our limits are approaching the 1~TeV regime. 
In most of the searches, the measurement sensitivity is limited by the size of the data sample. We therefore expect further improvements of our search as the data size increases.

\section*{Acknowledgments}
We thank CERN for the successful operation of the LHC. We acknowledge the enduring support of the
participating institutions and funding agencies, and the contributions of the collaborators
in the CMS experiment. Special thanks to the small team of people directly committed to these
challenging searches.

\end{document}